\documentclass[twocolumn,showpacs,prc,superscriptaddress,proprietresses]{revtex4}
\usepackage[dvips]{graphicx}
\usepackage{amsmath,amssymb}
\usepackage{dcolumn}
\begin{document}
\title{Uncertainty relations in terms of Tsallis entropy}
\author{Grzegorz Wilk}
\email{wilk@fuw.edu.pl} \affiliation{The Andrzej So{\l}tan
Institute for Nuclear Studies, Ho\.{z}a 69, 00681, Warsaw, Poland}
\author{Zbigniew W\l odarczyk}
\email{wlod@pu.kielce.pl} \affiliation{Institute of Physics,
                Jan Kochanowski University,  \'Swi\c{e}tokrzyska 15,
                25-406 Kielce, Poland}
\date{\today}
\begin{abstract}
Uncertainty relations emerging from the Tsallis entropy are
derived and discussed. In particular we found a positively defined
function that saturates the so called entropic inequalities for
entropies characterizing the physical states under consideration.
\end{abstract}

\maketitle

\section{\label{sec:I}Introduction}

It is known that the usual uncertainty relations, as given by the
Heinserberg uncertainty principle, $\Delta x \Delta p \ge
\frac{h}{4\pi}$ (which are based on standard deviations, $\Delta
x$ and $\Delta p$) frequently encounter serious difficulties
\cite{IBB_basic,PP}. The best examples are the cases of
probability distributions for which these deviations lose their
usefulness (being, for example, divergent). It was therefore
argued that one should base the formulation of these relations on
the information theory approach (see, for example, discussion in
\cite{IBB_basic} and references therein). In this way one avoids
the above mentioned problems. The price to be paid is, however,
the fact that the information theory approach depends on the type
of information measure used, which amounts to dependence on the
type of information entropy defining this measure. Examples of
Shannon, R\'enyi and Tsallis information entropies used for this
purpose are presented, for example, in \cite{IBB},
\cite{BR,ZV1,ZV2,L} and \cite{R,PP}, respectively (for more
information see references therein).

Let us notice that the entropic inequality relations involve sums
of entropies and are quite different from the standard uncertainty
relations. In standard uncertainty relations the product $\Delta x
\Delta p$ is strictly determined (i.e., $\Delta p$ is given by
$\Delta x$ and vice versa) for a given distribution function and
cannot take any values as will be the case further on below. The
uncertainty relation such as $\Delta x \Delta p \ge h/4\pi$  is
not a statement about the accuracy of our measuring instruments.
In contrast, entropic uncertainty relations do depend on the
accuracy of the measurement as they explicitly contain the area of
the phase space determined by the resolution of the measuring
instruments. In this paper we shall revisit, in Section
\ref{sec:II}, uncertainty relations emerging from Tsallis entropy
\cite{T} and discuss them in detail.  Our main result is present
in Section \ref{sec:III} in which we derive the new entropy
saturation function. Section \ref{sec:IV} contains our summary.\\

\section{\label{sec:II}Uncertainty relations emerging from
Tsallis entropy}

Let us define probability distributions associated with the
measurements of momentum ($p$) and position ($x$) of a quantum
particle in a pure state as
\begin{equation}
p_k = \int_{k \delta p}^{(k+1)\delta p}\! \! \! \! \! dp
\left|\tilde{\psi}(p)\right|^2;\quad x_l = \int_{l \delta
x}^{(l+1)\delta x}\!\!\! \! \! dx \left|\psi(x)\right|^2,
\label{eq:px}
\end{equation}
where indices $k$ and $l$ run from $0$ to $\pm \infty$ and the
Fourier transform is defined with the physical normalization ($h$
is Planck constant), i.e.,
\begin{equation}
\tilde{\psi}(p) = \frac{1}{\sqrt{h}} \int dx \exp\left( -
\frac{2\pi}{h} i px\right) \psi(x). \label{eq:Ft}
\end{equation}
 From the probability distributions $p_k$ and $x_l$ we may
construct the corresponding Tsallis entropies \cite{T}, which
measure the uncertainties in momentum and position spaces:
\begin{equation}
H^{(p)}_{\alpha} = \frac{\sum_k p^{\alpha}_k - 1}{1-\alpha};\qquad
H^{(x)}_{\beta} = \frac{\sum_l x^{\beta}_l - 1}{1-\beta}.
\label{eq:HpHx}
\end{equation}
In the respective limits of $(\alpha,\beta) \rightarrow 1$
entropies $H^{(r)}$ reduce to the Shannon entropy $(r=p,x)$:
\begin{equation}
S^{(r)} = - \sum_k r_k \ln r_k, \label{eq:S}
\end{equation}
for which the uncertainty relation has been derived long ago and
takes the form of a condition imposed on the sum of entropies
\cite{IBB},
\begin{equation}
S^{(p)} + S^{(x)} \ge - \ln \left( \frac{2\delta x \delta
p}{eh}\right) \label{eq:SpSx}
\end{equation}
(where $e$ is the basis of natural logarithm). The relation
(\ref{eq:SpSx}) reflects the fact that, although probability
distributions in Eq. (\ref{eq:px}) correspond to different
observables, nevertheless they describe the same quantum physical
state and therefore must be, in ge\-ne\-ral, correlated. Recently
Eq. (\ref{eq:SpSx}) has been generalized to the case of Renyi
entropies \cite{BR,ZV1,ZV2,L},
\begin{equation}
R^{(r)}_{\alpha} = \frac{1}{1-\alpha} \ln \left( \sum_k
r_k^{\alpha}\right), \label{eq:R}
\end{equation}
for which one gets \cite{BR} that
\begin{equation}
R^{(p)}_{\alpha} + R^{(x)}_{\beta} \ge -
\frac{1}{2}\left(\frac{\ln \alpha}{1-\alpha} + \frac{\ln
\beta}{1-\beta}\right) - \ln \left(\frac{2 \delta p \delta
x}{h}\right), \label{eq:RxRp}
\end{equation}
where parameters $\alpha$ and $\beta$ are assumed to be positive
and constrained by the relation
\begin{equation}
\frac{1}{\alpha} + \frac{1}{\beta} = 2 .\label{eq:condition}
\end{equation}

Let us now proceed to the case of nonextensive Tsallis entropy and
derive for it the corresponding entropic inequality. Our approach
differs from that already presented in \cite{R} in that we are
attempting from the very beginning to provide condition on the sum
of the corresponding $H^{(r)}_{\gamma}$ entropies (where $\gamma =
(\alpha, \beta)$, respectively). To do this we shall start from
the following Babenko-Beckner inequality relation \cite{BaBe},
\begin{equation}
\left[ (\delta p)^{1-\alpha}
\sum_kp^{\alpha}_k\right]^{\frac{1}{\alpha}}\!\!\! \leq \left(
\frac{2\alpha}{h}\right)^{\frac{-1}{2\alpha}}\!\!\!\left(
\frac{2\beta}{h}\right)^{\frac{1}{2\beta}}\!\!\!\left[ (\delta
x)^{1-\beta} \sum_lx_l^{\beta} \right]^{\frac{1}{\beta}}
\label{eq:inequality}
\end{equation}
which has been also used in \cite{BR} (cf., Eq. (21) there).
Parameters $\alpha$ and $\beta$ satisfy condition
(\ref{eq:condition}) and we shall assume at this moment that
$\alpha > \beta$. Notice that (\ref{eq:condition}) means that the
effects of nonextensivity in $x$ and $p$ spaces, as measured by
$\alpha$ and $\beta$, cannot be identical ($\alpha =\beta$ only
for $\alpha =1$ and $\beta = 1$, i.e., in the case  of the Shannon
entropy). The more general case of independent indices has been
recently discussed in \cite{ZV2} but we shall not comment on it
here. The inequality (\ref{eq:inequality}) can be rewritten as
\begin{equation}
 - \left( \sum_k p_k^{\alpha}
\right)^{\frac{1}{\alpha}} \ge - \eta(\alpha, \beta) \left( \sum_l
x_l^{\beta} \right)^{\frac{1}{\beta}}, \label{eq:ineq}
\end{equation}
where
\begin{equation}
\eta (\alpha, \beta) = \left(
\frac{\beta}{\alpha}\right)^{\frac{1}{2\alpha}} \left(
\frac{2\beta \delta x \delta p}{h} \right)^{\frac{\alpha
-1}{\alpha}}, \label{eq:etanew}
\end{equation}
or as
\begin{equation}
-1 + \frac{\alpha -1}{\alpha}A^{(p)}_{\alpha} \ge
-\eta(\alpha,\beta) + \eta(\alpha,\beta)\frac{\beta -
1}{\beta}A^{(x)}_{\beta} , \label{eq:111}
\end{equation}
where we have used the first order homogenous entropy defined as
(as before, $r=(p,x)$):
\begin{equation}
A^{(r)}_{\alpha} = \frac{\alpha}{\alpha - 1}\left[ 1 - \left( \sum
r^{\alpha}\right)^{\frac{1}{\alpha}}\right]  \label{eq:A}
\end{equation}
(it has been firstly introduced in \cite{A}, and then subsequently
given a complete characterization in \cite{BV}). By making use of
Eq. (\ref{eq:condition}) one can rewrite Eq. (\ref{eq:111}) in the
following way:
\begin{eqnarray}
A^{(p)}_{\alpha} &+& \eta(\alpha, \beta)A^{(x)}_{\beta} \ge
\frac{\alpha}{\alpha -1}[1-\eta(\alpha,\beta)] = \nonumber\\
& = & \frac{\alpha}{1-\alpha}\left[ \left( \frac{\beta}{\alpha}
\right)^{\frac{1}{2\alpha}} \left( \frac{2\beta}{h} \delta x
\delta p \right)^{\frac{\alpha -1}{\alpha}} - 1 \right].
\label{eq:222}
\end{eqnarray}

\begin{figure}[t]
\begin{center}
   \includegraphics[width=8.0cm]{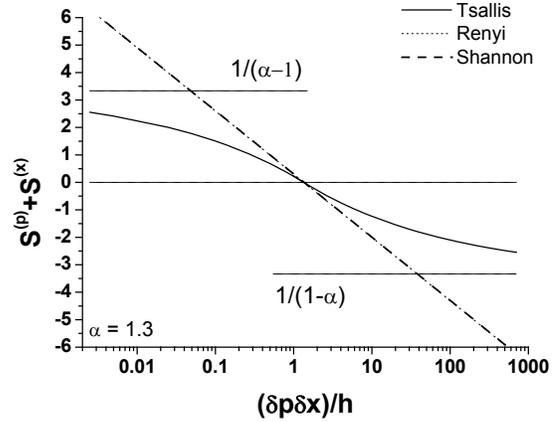}
   \vspace{-4.5cm}
   \caption{Example of the dependence (for $\alpha=1.3$) of limitations of the
   sum of entropies on the size of cell in phase space, $dxdp/h$.
   Results for Renyi and Shannon entropies are practically indistinguishable
   (to expose both the large and small values we used the
   linear-log scale here, in this case for both entropies one gets straight
   lines). Horizontal lines indicate the corresponding bounds for limitation
   imposed on Tsallis entropies for which $1/(1-\alpha) < \inf\{ H^{(x)}+H^{(p)}\} <
   1/(\alpha-1)$.}
   \label{fig1}
\end{center}
\vspace{-5mm}
\end{figure}

Further discussion depends on whether defined by Eq.
(\ref{eq:etanew}) coefficient $\eta(\alpha,\beta)$ is smaller or
greater than unity. In the first case
\begin{equation}
\eta(\alpha,\beta) \leq 1\quad {\rm or}\quad \delta x \delta p <
\frac{1}{2\beta}\left( \frac{\alpha}{\beta}
\right)^{\frac{1}{2(\alpha -1)}} h .\label{eq:etal1}
\end{equation}
We can now write the lhs of Eq. (\ref{eq:222}) as
\begin{equation}
A^{(p)}_{\alpha} + A^{(x)}_{\beta} \ge A^{(p)}_{\alpha} +
\eta(\alpha,\beta) A^{(x)}_{\beta},
\end{equation}
make use of the fact that for $\alpha \ge 1$ and $\beta \leq 1$
one has (see Eq. (\ref{eq:condition})),
\begin{equation}
\alpha H_{\alpha} \ge A_{\alpha}\quad{\rm and}\quad
\frac{\beta}{2\beta -1 }H_{\beta} \ge A_{\beta} \label{eq:cond},
\end{equation}
and get finally  that
\begin{equation}
H^{(p)}_{\alpha} + H^{(x)}_{\beta} \ge \frac{1}{\alpha}\left[
A^{(p)}_{\alpha} + A^{(x)}_{\beta}\right] . \label{eq:HpHx}
\end{equation}
It means that in this case one has
\begin{eqnarray}
\!\!\!\!\!H^{(p)}_{\alpha} + H^{(x)}_{\beta}\!\! &\ge&\!\!
\frac{1}{1-\alpha}\left[ \left( \frac{\beta}{\alpha}
\right)^{\frac{1}{2\alpha}} \left( \frac{2\beta}{h} \delta x
\delta p \right)^{\frac{\alpha - 1}{\alpha}}\!\!\! -1 \!\right].
\label{eq:TxTp1}
\end{eqnarray}
In the second case, for $\eta(\alpha,\beta) > 1$, one gets
\begin{eqnarray}
H^{(p)}_{\alpha} + H^{(x)}_{\beta} \ge \frac{1}{\alpha -1}\left[
\left( \frac{\alpha}{\beta} \right)^{\frac{1}{2\alpha}} \left(
\frac{2\beta}{h}\delta x \delta p
\right)^{\frac{1-\alpha}{\alpha}}\!\!\! - 1\! \right] .
\label{eq:TxTp2}
\end{eqnarray}
Both results generalize Eq. (\ref{eq:SpSx}), the result for
Shannon entropy, to which they converge when $\alpha \rightarrow
1$ and $\beta \rightarrow 1$.

To extend the above results to the case of $\alpha < \beta$ one
should use the same Babenko-Beckner inequality \cite{BaBe} as in
Eq. (\ref{eq:inequality}) but with the role of $p$ and $x$
interchanged, $p \leftrightarrow x$).

The dependence of the limitations on the sum of entropies on the
size of cell in phase space is visualized in Fig. \ref{fig1}.\\

\section{\label{sec:III}The entropy saturation function}

The inequalities presented above are, so far, purely mathematical
in the sense that they allow for {\it negative} lower limits for
the corresponding sum of entropies. For example, the rhs of
equation Eq. (\ref{eq:SpSx}) is positive only for
\begin{equation}
\delta p\delta x \leq \frac{1}{2} eh. \label{eq:limS}
\end{equation}
Because the sum of entropies must be {\it non-negative} therefore
the condition provided by Eq. (\ref{eq:SpSx}) only works together
with Eq. (\ref{eq:limS}). The same reasoning can be performed for
the remain two entropies leading to the following additional
requirements for the products $\delta x\delta p$:
\begin{equation}
\delta x \delta p \leq \frac{1}{2}h \alpha^{\frac{1}{2(\alpha
-1)}}\beta^{\frac{1}{2(\beta - 1)}} . \label{eq:another}
\end{equation}

\begin{figure}[t]
\begin{center}
\vspace{-1cm}
   \includegraphics[width=8.0cm]{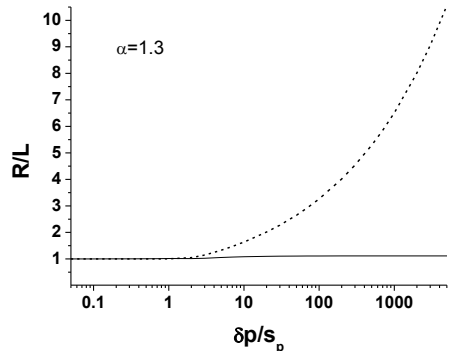}
   \vspace{-5.8cm}
   \caption{Examples of magnitude of the bias for the gaussian
            shape of probability densities, $\tilde{\rho}(p) =
            |\tilde{\psi}(p)|^ 2$ (calculated for $\alpha = 1.3$
            and $k = 0$). Broken line shows the ratio of the right
            to the left side of inequality (\ref{eq:D2a}) whereas
            the solid line shows the same for inequality
            (\ref{eq:D4a}) where we put $\delta t = \delta p/(s_p +
            \delta p)$.}
   \label{fig2}
\end{center}
\vspace{-5mm}
\end{figure}

The occurrence of negative values in the limitations of the sum of
entropies, $H^{(p)}_{\alpha} + H^{(x)}_{\beta}$, is the
consequence of the fact that for large values of $\delta x\delta
p/h$ we have $\eta(\alpha, \beta) > 1$. We shall now look at this
problem more closely. Evaluating $\eta(\alpha, \beta)$ we use the
integral form of Jensen's inequalities (which state that for
convex functions the values of the function at the average point
does not exceeds the average value of the function, the opposite
being true for concave functions \cite{Jensen}):
\begin{eqnarray}
\!\!\!\!\! \left[ \frac{1}{\delta p} \int^{(k+1)\delta p}_{k\delta
p} dp \tilde{\rho}(p)\right]^{\alpha} &\leq&  \frac{1}{\delta p}
\int^{(k+1)\delta p}_{k\delta p} dp \left[
\tilde{\rho}(p)\right]^{\alpha},
\label{eq:D2a}\\
\!\!\!\!\! \left[ \frac{1}{\delta x} \int^{(l+1)\delta x}_{l\delta
x} dx \rho (x)\right]^{\beta} &\ge&  \frac{1}{\delta x}
\int^{(l+1)\delta x}_{l\delta x} dx \left[ \rho(x)\right]^{\beta},
  \label{eq:D2b}
\end{eqnarray}
where the probability densities are
$\tilde{\rho}(p)=|\tilde{\psi}(p)|^2$ and $\rho (x) = |\psi
(x)|^2$ (cf. \cite{BR} for more details). It turns out that
differences between the left ($L$) and the right ($R$) hand sides
of inequalities (\ref{eq:D2a}) and (\ref{eq:D2b}) can be rather
substantial and can introduces serious bias to the results. Its
magnitude can be estimated using Taylor expansion: $E\left[
f\left(p_k\right)\right] \sim f\left[ E\left( p_k\right)\right] +
\frac{1}{2}f''\left[ E\left( p_k\right)\right] Var\left(
p_k\right)$ where $f(z) = z^{\alpha}$ \cite{DR}. However, this is
possible only when the functional form of probability $p_k$ is
known. In Fig. \ref{fig2} we show an example of the ration $R/L$
for inequality (\ref{eq:D2a}) calculated for a Gaussian shape of
$\tilde{\rho}(p) = |\tilde{\psi}(p)|^2$. The increase in
discrepancy is clearly visible. Instead of this, we shall now
demonstrate that the accuracy of Jensen's inequality can be
dramatically improved by a suitable change of variables. Namely,
we consider the following maps, which transform an infinite
interval to some fine interval, $r = (p,x) \in (-\infty, \infty)
\Longrightarrow t_r \in (-1,1)$:
\begin{equation}
t_r = \frac{r}{|r| + s_r} ,            \label{eq:D3}
\end{equation}
where $s_r$ is scale parameter such that $s_xs_p = h$. In new
variables the probability densities are given by
\begin{eqnarray}
\rho\left( t_x\right) &=& \rho\left[\ x\left(t_x\right) \right]\left|
\frac{dx}{dt_x}\right| = \rho (x) \frac{s_x}{\left( \left| t_x\right|
- 1\right)^2}, \label{eq:tx}\\
\tilde{\rho}\left( t_p\right) &=& \tilde{\rho}\left[\
p\left(t_p\right) \right]\left| \frac{dp}{dt_p}\right| =
\tilde{\rho} (p) \frac{s_p}{\left( \left| t_p\right| -
1\right)^2}. \label{eq:tp}
\end{eqnarray}
Using these new variables in analogous way as in Eqs.
(\ref{eq:D2a}) and (\ref{eq:D2b}), one can now write the following
inequalities:
\begin{eqnarray}
\!\!\!\!\!\!\!\!\!\!\left[ \frac{1}{\delta t_p} \int^{(k+1)\delta
t_p}_{k\delta t_p} \!\!\!\!\! \tilde{\rho}\left( t_p\right)  dt_p
\right]^{\alpha}\!\!\!\!\! &\leq&\!\! \frac{1}{\delta t_p}
\int^{(k+1)\delta t_p}_{k\delta t_p}\!\!\!\!\! \left[
\tilde{\rho}\left( t_p\right)\right]^{\alpha}
dt_p ,\label{eq:D4a}\\
\!\!\!\!\!\!\!\!\!\! \left[ \frac{1}{\delta t_x} \int^{(l+1)\delta
t_x}_{l\delta t_x} \!\!\!\!\!\rho \left( t_x\right) dt_x
\right]^{\beta}\!\!\!\!\! &\ge&\!\! \frac{1}{\delta t_x}
\int^{(l+1)\delta t_x}_{l\delta t_x}\!\!\!\!\! \left[ \rho\left(
t_x\right)\right]^{\beta}dt_x. \label{eq:D4b}
\end{eqnarray}
The ratio  $R/L$ for inequality (\ref{eq:D4a}) calculated for a
Gaussian shape of $\tilde{\rho}(p) = |\tilde{\psi}(p)|^2$ is
shown in Fig. \ref{fig2} and, as one can see, grows very weakly
with the bin size $\delta p$.

Establishing this finding, let us now proceed to a calculation of
the corresponding entropic inequalities using new variables. The
probabilities corresponding to (\ref{eq:px}) are now:
\begin{equation}
p'_k = \int^{(k+1)\delta t_p}_{k\delta t_p}\!\!\!\!\!
\tilde{\rho}\left(t_p\right)dt_p,\quad x'_l = \int^{(l+1)\delta
t_x}_{l\delta t_x}\!\!\!\!\! \rho\left(t_x\right)dt_x
\label{eq:D5} .
\end{equation}
(notation is such that primed quantities correspond to using the
new variable $t_r$ and non-primed ones to the standard variable
$r=(x,p)$). Whereas before, in variables $(x,p)$, $k$ and $l$ were
varying from $0$ to $\pm \infty$, now $k \in \left(0, \pm
k_{max}\right)$ and $l \in \left(0, \pm l_{max}\right)$ where
$k_{max} \delta t_p = 1$ and $l_{max} \delta t_x = 1$. For these
probabilities we get the following equivalent of Eq.
(\ref{eq:ineq}),
\begin{equation}
 - \left( \sum_k {p'}_k^{\alpha}
\right)^{\frac{1}{\alpha}} \ge - \eta(\alpha, \beta) \left( \sum_l
{x'}_l^{\beta} \right)^{\frac{1}{\beta}}, \label{eq:D6}
\end{equation}
where now
\begin{equation}
\eta(\alpha, \beta) = \left(
\frac{\beta}{\alpha}\right)^{\frac{1}{2\alpha}} \left( 2 \beta
\delta t_x \delta t_p \right)^{\frac{\alpha -1}{\alpha}}.
\label{eq:D7}
\end{equation}
Notice that now $\eta (\alpha, \beta)  \leq 1$ {\it always}, this
means that we shall no more encounter problems with negative
values for the limits of the sum of entropies.

To be more specific, notice that for entropies
${H'}^{(p)}_{\alpha} = \left[ 1 -
\sum\left(p'_k\right)^{\alpha}\right]/(\alpha - 1)$ and
${H'}^{(x)}_{\beta} = \left[ 1 -
\sum\left(x'_l\right)^{\beta}\right]/(\beta - 1)$ we have that
\begin{eqnarray}
{H'}^{(p)}_{\alpha} + {H'}^{(x)}_{\beta} \ge \frac{1}{\alpha
-1}\left[ 1 - \left(
\frac{\beta}{\alpha}\right)^{\frac{1}{2\alpha}}\!\!\! \left(2\beta
\delta t_x \delta t_p\right)^{\frac{\alpha -1}{\alpha}}\right]
\label{eq:D8}.
\end{eqnarray}
Putting $\alpha \rightarrow 1$ and proceeding to Shannon entropy
 one gets that (in bits)
\begin{equation}
{S'}^{(p)} + {S'}^{(x)} \ge \left[ \ln \left(\frac{1}{\delta t_x
\delta t_p}\right) + 1\right]\frac{1}{\ln 2} - 1 . \label{eq:D9}
\end{equation}
It is interesting to note that for the uniform distribution in the
variable $t_r \in (-1,1)$ one has
$$ \left[ \ln \left(\frac{1}{\delta t_x
\delta t_p}\right)\right]\frac{1}{\ln 2} + 2$$ bits of information
(the number of bins are $2/\delta t_x$ and $2/\delta t_p$). The
interval of variability of ${S'}^{(p)} + {S'}^{(x)}$ is narrow and
equals $ 3 - 1/ln 2 \simeq 1.557$ bits (this is the difference
between the maximal and minimal limitations).

Let us notice at this point that, whereas inequalities
(\ref{eq:SpSx}), (\ref{eq:RxRp}), (\ref{eq:TxTp1}) and
(\ref{eq:TxTp2}) are for the fixed values of intervals $\delta x$
and $\delta p$, the inequality (\ref{eq:D8}) is for the fixed
values of intervals $\delta t_x$ and $\delta t_p$. Formal
recalculation of these intervals results in their dependence on
$k$ and $l$, they are not fixed anymore but their values change in
the following way: for $\delta t_r = const$ one has
\begin{equation}
\delta r = \int^{(k+1)\delta t_r}_{k \delta
t_r}\!\!\!\!\!\!\!\!\!\! \frac{s_r}{\left(\left| t_r\right| - 1
\right)^2} dt_r = \frac{ s_r \delta t_r}{\left(1 - |k|\delta
t_r\right)\left[ 1 - |k+1|\delta t_r \right]}, \label{eq:D10}
\end{equation}
whereas for $\delta r = const$ one has
\begin{equation}
\delta t_r = \int^{(k+1)\delta r}_{k \delta r}\!\!\!\!\!\!\!\!\!\!
\frac{s_r}{\left(s_r + |r|\right)^2} dr = \frac{ s_r \delta
r}{\left(s_r + |k|\delta r\right)\left[ s_r + |k+1|\delta r
\right]} . \label{eq:D11}
\end{equation}
Notice that because Eq. (\ref{eq:D3}) is an odd function of $r$
and has rotational symmetry with respect to the origin, one has
exactly the same intervals $\delta r$ and $\delta t_r$ for the
negative values of $k$, $k = - \kappa$, and for its positive
values, $k = \kappa -1$, where $\kappa = 1,2,3,\dots$.

The natural question is then in what way, for some given fixed
intervals $\delta r = (\delta x, \delta p)$, one should choose
intervals $\delta t_r = \left(\delta t_x,\delta t_p\right)$ in
inequality (\ref{eq:D8}). If we take the maximal values of
intervals $\delta t_r$ (corresponding to $k=0 $ or $k = -1$) and
make use of the fact that now
\begin{equation}
\delta t_x\delta t_p = \frac{\delta x}{s_x + \delta x}\frac{\delta
p}{s_p + \delta p} \leq \frac{\delta x\delta p}{h + \delta x
\delta p}, \label{eq:D12}
\end{equation}
then we obtain that the right-hand-side of inequality
(\ref{eq:D8}) will be limited by
\begin{eqnarray}
\frac{1}{\alpha -1}&&\!\!\!\!\! \left[ 1 - \left(
\frac{\beta}{\alpha}\right)^{\frac{1}{2\alpha}} \left(2\beta
\delta t_x \delta t_p\right)^{\frac{\alpha -1}{\alpha}}\right] \ge
\nonumber\\
 && \!\!\!\!\!\! \!\!\!\!\! \!\!\!\ge \frac{1}{\alpha -1}\left[ 1 - \left(
\frac{\beta}{\alpha}\right)^{\frac{1}{2\alpha}} \left(2\beta
\frac{\delta x \delta p}{h + \delta x\delta
p}\right)^{\frac{\alpha -1}{\alpha}}\right]. \label{eq:D13}
\end{eqnarray}
Actually, taking exactly the results of (\ref{eq:D10}) and
(\ref{eq:D11}) we would obtain equality, not inequality in Eq.
(\ref{eq:D13}). However, in such case one would not have at the
same time $\delta x = const$  and $\delta t_x = const$ (or $\delta
p = const$ and $\delta t_p = const$). Choosing intervals
corresponding to $k=0$ or $k = -1$ (for which we have maximal
interval $\delta t_p$ equal to $\delta t_p=\delta p /\left(s_p +
\delta p\right)$ or, equivalently, minimal interval $\delta p$
equal to $\delta p = s_p \delta t_p/\left(1 - \delta t_p\right)$)
we can see that for each $p'_{k'}$ (given by Eq. (\ref{eq:D5})) we
have $p_k$ (given by Eq. (\ref{eq:px})), which satisfies the
inequality $p_k \leq {p'}_{k'}$ and for $\alpha > 1$ we have
$\left(p_k\right)^{\alpha} \leq \left( {p'}_{k'}\right)^{\alpha}$
(see \footnote{The probability densities are transformed according
to Eqs. (\ref{eq:tx}) and (\ref{eq:tp}). For exact transformation
of bins (given by Eqs. (\ref{eq:D10}) and (\ref{eq:D11})) we have
the same probabilities, $p_k\left(\bar{r}\right)
=p'_k\left(\bar{t}_r\right)$. Choosing maximal interval $\delta
t_p$, for the same probability densities we have inequality
$p_k\left(\bar{r}\right) \leq p'_{k'}\left(\bar{t}_r\right)$.
Notice now that the number of bins in both cases is different and
that probabilities in this inequality are not for the same bin
number $k$ but for the corresponding position in variables $r$ and
$t_r$.}). However, because the number of bins in both cases is
different, there will be some $p_k$ left for which there will be
no $p'_{k'}$ assigned. Nevertheless, one can construct some new
$p_k$'s by performing division of $p'_{k'}$. Preserving always the
normalization, i.e., assuming that $\sum_k p_k = \sum_{k'}
{p'}_{k'} = 1$ one has that \footnote{For the condition $\sum_k
p_k = \sum_k {p'}_{k} = 1$ increasing the number of divisions
leads, for $\alpha > 1$, to decreasing of $\sum \left(
p_k\right)^{\alpha}$ (and to its increasing for $\alpha < 1$). Let
us notice that $\sum^n_k {p'}^{\alpha}_{k} = \sum^{n-1}_k
{p'}^{\alpha}_k + {p'}^{\alpha}_n$. Putting therefore $p_k =
p'_{k}$ for $k = 0,1,2,\dots, n-1$ and $p_n + p_{n+1} = p'_n$ one
gets, for $\alpha
> 1$, that $p'^{\alpha}_n = \left( p_n + p_{n+1}\right)^{\alpha}
\ge p_n^{\alpha} + p^{\alpha}_{n + 1}$, which leads to $\sum_k^n
{p'}^{\alpha}_{k} \ge \sum_k^{n+1} p^{\alpha}_k$. Repeating this
procedure of dividing the ${p'}_{k}$ (and possibly also dividing
again $p_k$) one gets that $\sum_k^n{p'}^{\alpha}_{k} \ge \sum^m_k
p^{\alpha}_k$, where $m > n$. Actually, this inequality is true
also for $m \rightarrow \infty $. To see it, let us consider
$\varepsilon_m = \sum^{\infty}_{|k|=m+1}\tilde{p}_k$ and
$\varepsilon_m^{(\alpha)} =
\sum^{\infty}_{|k|=m+1}\tilde{p}^{\alpha}_k$ for probabilities
$\tilde{p}_k$ with normalization $\sum^{\infty}_{|k|}\tilde{p}_k =
1$. We can write $\sum^m_{|k|} p^{\alpha}_k =
\left(\sum_{|k|}^{\infty}\tilde{p}^{\alpha}_k -
\varepsilon^{(\alpha)}_m\right)/\left(1 -
\varepsilon_m\right)^{\alpha}$. Because for the physically
motivated probability distributions both $\lim_{m \to \infty}
\varepsilon_m = 0$ and  $\lim_{m \to \infty}
\varepsilon_m^{(\alpha)} = 0$, we get that $\lim_{m \to
\infty}\sum^m_{|k|} p^{\alpha}_k = \sum^{\infty}_{|k|}
\tilde{p}^{\alpha}_k$.}
\begin{equation}
\sum \left(p_k\right)^{\alpha} \leq \sum \left(
{p'}_{k'}\right)^{\alpha} . \label{eq:D14}
\end{equation}
We have then for entropies  $H_{\alpha} = \left[ 1 - \sum
\left(p_k\right)^{\alpha}\right]/(\alpha - 1)$ and $H'_{\alpha} =
\left[ 1 - \sum\left(p'_{k'}\right)^{\alpha}\right]/(\alpha - 1)$
the inequality that $H_{\alpha} \ge H'_{\alpha}$. Analogously,
repeating the above procedure for probabilities $x_l$ and
$x'_{l'}$ one gets that $H_{\beta} \ge H'_{\beta}$ (where now,
according to (\ref{eq:condition}), $\beta < 1$). The limitation
for the left-hand-side of inequality (\ref{eq:D8}) is then
\begin{equation}
H^{(p)}_{\alpha} + H^{(x)}_{\beta} \ge {H'}^{(p)}_{\alpha} +
{H'}^{(x)}_{\beta}. \label{eq:D15}
\end{equation}

Finally, for the Tsallis entropy we can write:
\begin{equation}
H^{(p)}_{\alpha} + H^{(x)}_{\beta} \ge \frac{1}{\alpha -1}\left[ 1
- \left( \frac{\beta}{\alpha}\right)^{\frac{1}{2\alpha}}
\left(2\beta \frac{\delta x \delta p}{h + \delta x\delta
p}\right)^{\frac{\alpha -1}{\alpha}}\right]. \label{eq:D16}
\end{equation}
For $\delta x\delta p/h << 1$ we recover the previous result given
by Eq. (\ref{eq:TxTp1}) whereas in the limit of $\delta x\delta
p/h \rightarrow \infty$ we have
\begin{equation}
H^{(p)}_{\alpha} + H^{(x)}_{\beta} \longrightarrow \frac{1}{\alpha
- 1}\left[ 1 -
(2\alpha)^{-\frac{1}{2\alpha}}(2\beta)^{\frac{1}{2\beta}}\right]
> 0. \label{eq:D17}
\end{equation}
Notice that now the limit is always positive.

\begin{figure}[t]
\vspace{-1cm} \hspace{-1.5cm}
   \includegraphics[width=10.0cm]{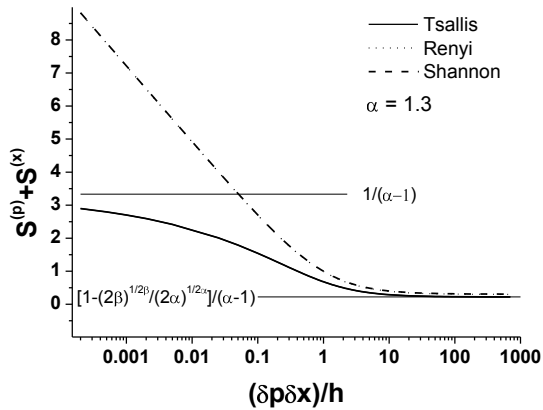}
   \vspace{-7.5cm}
   \caption{Example of the dependence (for $\alpha = 1.3$) of
            limitations of the sum of entropies on the size of cell
            in phase space, $\delta x\delta p/h$. Results for Renyi
            and Shannon entropies are practically indistinguishable.
            Horizontal lines indicate corresponding limits for the
            sum of Tsallis entropies for which
            $1/(\alpha - 1) \leq \inf \{H^{(p)}_{\alpha}
            + H^{(x)}_{\beta}\} \leq
            [1 - (2\alpha)^{-1/2\alpha}(2\beta)^{1/2\beta}]/(\alpha -
            1)$.}
   \label{fig3}
\vspace{-5mm}
\end{figure}

In the limit $\alpha \rightarrow 1$ we get a limitation for
Shannon entropy, which now reads
\begin{equation}
S^{(p)} + S^{(x)} \geq - \ln \left( \frac{2}{e}\frac{\delta x
\delta p}{h + \delta x \delta p}\right). \label{eq:D18}
\end{equation}
For large intervals, i.e, for $\delta x\delta p/h \rightarrow
\infty$, one gets $S^{(p)} + S^{(x)} \rightarrow 1 - \ln 2$
\footnote{Actually this bound was first conjectured by Hirschman
\cite{H} and proven by Beckner \cite{Beckner}. For Renyi entropy
it was derived in \cite{HS} and reads: $R^{(p)}_{\alpha} +
R^{(x)}_{\beta} \geq \frac{1}{2}\left[ \ln \alpha/(\alpha - 1) +
\ln \beta /(\beta - 1)\right] - \ln 2$. Our Eq. (\ref{eq:D17})
extends therefore Hirschman uncertainty to Tsallis entropy.}. It
should be noticed that this new inequality (\ref{eq:D18}) for
Shannon entropy is stronger (for all values of interval $\delta p
\delta x$) than the previous limitation (\ref{eq:SpSx}) derived in
\cite{IBB}. The new dependencies of limitations on different
entropies on the size of the phase space cell $\delta x\delta p/h$
are displayed in Fig. \ref{fig3} \footnote{To get a feeling of
difference between limitations represented by Eqs. (\ref{eq:SpSx})
and (\ref{eq:D18}) let us consider, as example, gaussian
probability densities (corresponding to the gaussian
wave-function) with dispersion equal unity, for which we can
evaluate $S^{(p)}$ and $S^{(x)}$ using definition (\ref{eq:S}).
For $\delta x/s_x = \delta p/s_p = \sqrt{2}$ the sum of entropies
is equal to $S^{(p)} + S^{(x)} \cong 1.76$. The lower limit
provided in this case by Eq. (\ref{eq:D18}) is $S^{(p)} + S^{(x)}
\ge 0.712$, which is much stringent than the corresponding limit
$S^{(p)} + S^{(x)} \ge - 0.386 $ provided by Eq. (\ref{eq:SpSx})
(notice that it leads to negative value of the sum of entropies).}
.

\section{\label{sec:IV}Summary}

We have derived uncertainty relations based on Tsallis entropy. We
have also found a positively defined function that saturates the
so called entropic inequalities for entropies characterizing
physical states under consideration, cf. Eq. (\ref{eq:D16}). In
case of Shannon entropy (Eq. (\ref{eq:D18})) the limit provided is
more stringent than the previously derived. Formally, our results
show that changing $\delta p \delta x/h$ to $\delta p \delta x/(h
+ \delta p \delta x)$ one avoids (in all cases: Shannon, Renyi and
Tsallis entropies) the appearance of unphysical negative values in
the entropy bounds.

Let us close with the remark that in some applications of the
nonextensive statistics the nonextensivity parameter $q$
(corresponding to $\alpha$ and $\beta$ here) describes intrinsic
fluctuations existing in the physical system under consideration
\cite{qWW}. This raises an interesting question of the possible
existence of such relations also in the applications mentioned
above. In particular there still remains the question of whether
our results will survive the other choices of inequality used in
(\ref{eq:inequality}) and/or in the case of independent indices
($\alpha$, $\beta$) as discussed in \cite{ZV2}. We plan to address
this point elsewhere.\\

\begin{acknowledgments}
Partial support (GW) from the Ministry of Science and Higher
Education under contract 1P03B02230 is acknowledged.
\end{acknowledgments}


\end{document}